\documentstyle[preprint,aps,eqsecnum]{revtex}

\begin{document}
\draft

\title{Replica Field Theory for a Polymer in Random Media}
\author{Yadin Y. Goldschmidt}
\address{Department of Physics and Astronomy\\University of Pittsburgh\\%
Pittsburgh, PA 15260}
\date{Aug. 12, 1999}
\maketitle
\begin{abstract}
In this paper we revisit the problem of a (non self-avoiding) polymer chain in
a random medium which was previously investigated by Edwards and Muthukumar
(EM) \cite{EM}. As noticed by Cates and Ball (CB)\cite{cates} there is a
discrepancy between the predictions of the replica calculation of EM and the
expectation that in an infinite medium the quenched and annealed results
should coincide (for a chain that is free to move) and a long polymer should
always collapse. CB argued that only in a finite volume one might see a
``localization transition'' (or crossover) from a stretched to a collapsed
chain in three spatial dimensions. Here we carry out the replica calculation
in the presence of an additional confining harmonic potential that mimics the
effect of a finite volume. Using a variational scheme with five variational
parameters we derive analytically for $d<4$ the result $R\sim\left(  g\left|
\ln\mu\right|  \right)^{-1/(4-d)}\sim\left(  g \ln V \right)
^{-1/(4-d)}$, where $R$ is the radius of gyration, $g$ is the strength of the 
disorder, $\mu$ is the spring constant associated with the confining potential 
and $V$ is the associated effective volume of the system. Thus the EM\ result 
is recovered with their constant replaced by $ \ln V $ as argued by CB. We see 
that in the strict infinite volume limit the polymer always collapses, but for 
finite volume a transition from a stretched to a collapsed form might be 
observed as a function of the strength of the disorder. For $d<2$ and for large
$V>V^{\prime}\sim\exp(g^{2/(2-d)}L^{(4-d)/(2-d)})$ the annealed results are
recovered and $R\sim(Lg)^{1/(d-2)}$, where $L$ is the length of the polymer.
Hence the polymer also collapses in the large $L$ limit. The 1-step replica
symmetry breaking solution is crucial for obtaining the above results.
\end{abstract}
\newpage

\section{Introduction}

There has been much interest in recent years in the properties of polymer
chains in a quenched random environment \cite{EM,cates,book,BM,vilgis}. This
problem is directly related to that of a quantum particle in a random medium
\cite{yygold,chen} and to that of a flux line in a type II superconductor in
the presence of random columnar defects \cite{vinokur,yadin} as will be made
clear below. Thus its general application makes it important for a variety of
physical situations.

The quantities of interest for the polymer problem are the free energy and the
radius of gyration of a chain in a quenched white-noise potential. Here we
consider only the case of a non self-avoiding chain. Cates and Ball
\cite{cates} gave a beautiful intuitive argument to the effect that a Gaussian
chain situated in an infinite random medium is always collapsed in the
long-chain limit. Their argument goes as follows: Consider a white-noise
random potential $v(x)$ of zero mean whose probability distribution at each 
site is:
\begin{equation}
P(v(x))\ \propto\ g^{-1/2}\ \exp(-v^{2}/2g).\label{random}%
\end{equation}

If we now coarse-grain the medium and denote by $\overline{v}$ the average
value of the potential over some region of volume $a$, Then the coarse-grained
potential will have the distribution
\begin{equation}
P_{a}(\overline{v})\ \propto\ (g/a)^{-1/2}\ \exp(-a\overline{v}%
^{2}/2g).\label{coarse}%
\end{equation}

Consider a polymer chain situated in the random potential, and assume that it
shrinks into a volume $a$ corresponding to a place where the mean potential
$\overline{v}$ takes on a lower value than usual. In this situation the free
energy of the chain is crudely estimated to be (neglecting all numerical
factors):
\begin{equation}
F(a,\overline{v})=L/R^{2}+L\overline{v}+a\overline{v}^{2}/2g.\label{fe}%
\end{equation}
Here $L$ is the length of the chain (number of monomers), $R$ is the radius of
gyration (or end to end distance) and the volume $a$ is related to $R$ via
$a=R^{d}$ in $d$-spatial dimensions. The first term on the r.h.s. is an
estimate of the free energy of a long chain confined to a region of size $R$
in the absence of an external potential (see e.g. \cite{de-gennes}, Eq. I.12).
The second term is just the potential energy of the chain in the random
potential of strength $\overline{v}$. The third term arises from the chance of
incurring a random potential of strength $\overline{v}$. The quantity $\ln
P(\overline{v})$ gives an associated effective entropy for the system.
Minimizing this free energy over both $\overline{v}$ and $a$ determines the
lowest free energy configuration. Minimizing with respect to $\overline{v}$
yields $\overline{v}=-Lg/a$. Substituting in $F$ gives:
\begin{equation}
F(R)=\frac L{R^{2}}-\frac{L^{2}g}{2R^{d}}.\label{fran}%
\end{equation}
This shows that for any $d\geq2$, $F\rightarrow-\infty$ as $R\rightarrow0 $.
Thus the mean size of the chain is zero, or in the presence of a cutoff of a
size of one monomer,
\begin{equation}
R\ \sim1,\ \ \ d\geq2.\label{Rg2}%
\end{equation}
For $d<2$, the free energy has a minimum for
\begin{equation}
R\sim(Lg)^{1/(d-2)}\ \ d<2\ ,\label{Rl2}%
\end{equation}
which in the long chain limit ($L\geq1/g$) cuts off again at $R\ \sim1$. These
results are the same as those for the case of an annealed potential that is
able to adjust locally to lower the free energy of the system. The reason is
that for an infinite system containing a finite (even though long) chain,
space can be divided into regions containing different realizations of the
potential, and the chain can sample all of these to find an environment
arbitrarily similar to that which would occur in the annealed situation.

These results stand in contrast to the replica calculation of Edwards and
Muthukumar (EM) \cite{EM}, who found that for a long chain
\begin{equation}
R\sim g^{-1/(4-d)},\ \ \ \ d<4\label{REM}%
\end{equation}
when $g^{2/(4-d)}L\rightarrow\infty$ \ , whereas $R\sim L^{1/2}$ when
$g^{2/(4-d)}L\rightarrow0$. Note that the result (\ref{REM}) is independent of
$L$ as opposed to Eq. (\ref{Rl2}). To reconcile the two apparently different
results, Cates and Ball argue that the quenched case is different from the
annealed case only for the case when the medium has a \textit{finite }volume
$V$. In a finite box, arbitrarily deep potential minima are not present.
Instead the most negative $\overline{v}$ averaged over a region of volume
$a\ \ll V$ occupied by the chain, is approximately (keeping only leading terms
in the volume $V$) given by solving the equation (the l.h.s. of which
represents the area under the tail of the distribution)
\begin{equation}
\int_{-\infty}^{\overline{v}}dy\ P_{a}(y)\simeq\frac{a}{V}\ ,\label{ieq}%
\end{equation}
which yields
\begin{equation}
\overline{v}=-\sqrt{\frac{g\ln V}{a}}.\label{vmin}%
\end{equation}
This expression when plugged into Eq. (\ref{fe}) leads to (Note that the last
term in (\ref{fe}) just becomes a constant independent of $R$)
\begin{equation}
F(R)=\frac{L}{R^{2}}-L\sqrt{\frac{g\ln V}{R^{d}}}.\label{FQ}%
\end{equation}
When this free energy is minimized with respect to $R$ \ it gives rise to
\begin{equation}
R\sim(g\ln V)^{-1/(4-d)},\ \ \ \ d<4\label{RV}%
\end{equation}
which agrees with Eq.(\ref{REM}) and also with simulations performed on a
chain in a random medium of a fixed finite volume \cite{BM}. However it is not
clear from this explanation why the replica calculation which has been done
for an infinite system \cite{EM} gives rise to the finite volume result. To
shed light on this question we will show in this paper that the reason for the
discrepancy is the fact that EM used a variational calculation which relies on
a single variational parameter. We show specifically that the single parameter
variational solution is inconsistent.

What we will do first is, instead of considering a system in a finite volume
which is hard to solve, introduce an external harmonic potential (with a
spring constant $\mu$). Such an attractive potential has the effect of
confining the chain to a finite distance from the origin since the energy cost
to wonder far away from the origin of the potential is high. A system in a
harmonic potential is easier to solve than a system in a finite box. It also
corresponds directly to the problem of a flux line in a type II superconductor
where the cage potential felt by a flux line due to its neighboring flux lines
can be modeled by a harmonic potential (see below). In addition, we introduce
more variational parameters, three for the case of replica symmetric
parametrization and five for the case of replica symmetry breaking (RSB).
These extra parameters have physical significance as will be discussed below.
We will then use the replica method and the variational approximation to
tackle the problem and obtain the free energy and the radius of gyration. For
finite $\mu$ we find that $R$ is independent of $L$ (the chain length) and as
the disordered strength is increased from zero, $R$ is decreased from its
initial $\mu$-dominated value according to the relation $R\sim\left(  g\left|
\ln\mu\right|  \right)  ^{-1/(4-d)}$ (which agrees with Eq. (\ref{RV}) since
the effective volume available to a system in a harmonic potential is $\ln
V\sim\left|  \ln\mu\right|  $). 

On the other hand, if we try to take the
ultimate $\mu\rightarrow0$ limit (which is the case originally studied by EM),
the previous solution becomes invalid and the chain collapses for $d\geq2$.
For $d<2$ the annealed results are obtained in the $\mu\rightarrow0$ limit as
given above in equations (\ref{fran},\ref{Rg2},\ref{Rl2}). This occurs
specifically because of the extra variational parameters used beyond the
single variational parameter used by EM. We also demonstrate the importance of
RSB for obtaining the correct physical results. (The relevance of RSB to this
model was recognized by Haronska and Vilgis \cite{vilgis}, but unfortunately
their calculation still predicted a constant coefficient of proportionality in
the relation $R\sim\left(  cg\right)  ^{-1/(4-d)}$, that although differs from
the EM result does not contain the correct $\ln V$ dependence.)

To define the model of a polymer chain in a random potential plus a fixed
harmonic potential we use the Gaussian chain approximation to write:
\begin{equation}
H=\int_{0}^{L}du\ \left[  \frac{M}{2}\left(  \frac{\partial {\mathbf R}%
(u)}{\partial u}\right)  ^{2}+\frac{\mu}{2}{\mathbf R}^{2}(u)+V({\mathbf R}%
(u){\mathbf  )}\right]  ,\label{H}%
\end{equation}
were ${\mathbf  R}(u)$ is the $d$-dimensional position vector of the chain at
arc-length $u$ ($0\leq u\leq L$), $\mu$ governs the strength of the harmonic
potential and $V({\mathbf  R})$ is the random potential satisfying:
\begin{equation}
\left\langle V({\mathbf  R})\right\rangle =0,\ \ \ \left\langle V({\mathbf  R})%
V({\mathbf  R}^{\prime} )\right\rangle =g\ \delta^{(d)}({\mathbf  R-R}%
^{\prime}).\label{VV}%
\end{equation}
We can actually consider a wider class of random potential correlations
characterized by a function $f$:
\begin{equation}
\left\langle V({\mathbf  R})V({\mathbf  R}^{\prime} )\right\rangle
=g\ d\ f\left(  \frac{({\mathbf  R-R}^{\prime})^{2}}{d}\right)  ,\label{Vf}%
\end{equation}
where $f()$ is some given function. In Eq. (\ref{H}) we choose the units such
that $u$ is dimensionless and so $L$ is the length of the polymer in units of
the Khun bond step $b$. The ``mass'' $M$ is inversely proportional to $\beta
b^{2}$, where $\beta=1/k_{B}T$ $\ $(in $d$-dimensions $\beta M=d/b^{2}$). The
case ${\mathbf  R}(0)={\mathbf  R}(L)$ corresponds to a closed chain.

The partition sum is given by the functional integral
\begin{equation}
Z({\mathbf  R},{\mathbf  R}^{\prime},L,\beta)=\int_{{\mathbf  R}(0)={\mathbf  R}%
}^{{\mathbf  R}(L)={\mathbf  R}^{\prime}}{\mathcal D}{\mathbf  R}(u)\ \exp(-\beta
H).\label{Partition}%
\end{equation}
We further define a boundary-free partition sum (for a closed chain) by
\begin{equation}
Z(L,\beta)=\int d{\mathbf  R\ }Z({\mathbf  R},{\mathbf  R},L,\beta),\label{pf}%
\end{equation}
and the free energy is given by
\begin{equation}
\beta F=-\ln Z(L,\beta).\label{free}%
\end{equation}
The correlation function of interest is
\begin{equation}
C(\ell)=\frac1d\left\langle \left\langle ({\mathbf  R}(\ell){\mathbf  -R}%
(0))^{2}\right\rangle \right\rangle _{R}\ \ ,\label{R2}%
\end{equation}
where
\begin{equation}
1\ll\ell\ll L.\label{el}%
\end{equation}
The first average in Eq. (\ref{R2}) is the thermal one with a Boltzmann weight
exp(-$\beta H$) and the second average is over the realizations of the random
potential. For the range of $\ell$ given by Eq. (\ref{el}), the boundary
conditions on the chain, e.g. open or closed are not important for the
behavior of $C(\ell)$.

For the case of no disorder (i.e. $g=0$) the correlation function is given by
\begin{equation}
C_{0}(\ell)=\frac{1}{\beta\sqrt{M\mu}}\left(  1-\exp(-\ell\sqrt{\mu
/M})\right)  .\label{c0}%
\end{equation}
We see that in the limit $\mu\rightarrow0$, $C_{0}(\ell)\sim\ell\ /\ \beta M$
$\sim(b^{2}/d)\ \ell$, which corresponds to pure diffusion of the chain
(random walk). From the relation
\begin{equation}
\left\langle \left\langle {\mathbf  R}(0)^{2}\right\rangle \right\rangle
_{R}=\frac{d}{\beta\sqrt{M\mu}},
\end{equation}
we see that the polymer chain is confined to a volume of size $V$ satisfying
\begin{equation}
\ln V\sim\frac{d}{4}\left|  \ln\mu\right|  ,
\end{equation}
for small $\mu$.

The mapping of this problem to a vortex line in an harmonic cage potential and
random columnar defects is such that the arc-length $u$ corresponds to the
distance $z$ along the $c$-axis (assuming this is also the direction of the
magnetic field), $M\rightarrow\epsilon_{l}=\epsilon_{0}/\gamma^{2}$, which is
the line tension of the flux line and $\gamma^{2}=m_{z}/m_{\bot}$ is the mass
anisotropy. ${\mathbf  R}$ is a two dimensional vector in the $a-b$ plane of the
superconductor \cite{vinokur,yadin}. The harmonic potential plays an essential
role as a reasonable approximation to the cage potential that a vortex line
feels due to the repulsion by its neighbors. Thus $\mu\approx\epsilon
_{0}B/\Phi_{0}$ where $B$ is the magnetic field and $\Phi_{0}$ is the fluxoid.

There is also a mapping into the problem of a quantum particle in a random
potential + a harmonic potential. This mapping reads \cite{feynman,yygold}
\begin{equation}
\beta\rightarrow1/\hbar,\ \ \ L\rightarrow\beta\hbar,\label{map}%
\end{equation}
and $\rho({\mathbf  R},{\mathbf  R}^{\prime},\beta)= \ Z({\mathbf  R}%
,{\mathbf  R}^{\prime},L=\beta\hbar,\beta=1/\hbar)$ becomes the density matrix
of a quantum particle at inverse temperature $\beta$. The variable $u $
represents the Trotter (imaginary) time. In this case $M$ corresponds to the
mass of the particle.

\section{The variational calculation}

In order to average over the quenched random potential we use the replica
method. After introducing $n$-copies of the chain and averaging over the
random potential one obtains
\begin{equation}
\left\langle Z^{n}\right\rangle =\int{\mathcal D}{\mathbf  R}_{1}\cdots
{\mathcal D}{\mathbf  R}_{n}\exp(-\beta H_{n}),\label{zn}%
\end{equation}
with
\begin{eqnarray}
H_{n}  & =\int_{0}^{L}du\ \sum_{a=1}^{n}\left[  \frac M2\left(  \frac
{\partial{\mathbf  R}_{a}(u)}{\partial u}\right)  ^{2}+\frac\mu2{\mathbf  R}%
_{a}^{2}(u)\right] \nonumber\\
& -\frac{\beta g}2\int_{0}^{L}du\int_{0}^{L}du^{\prime}\ \sum_{ab}\delta
^{(d)}\left(  {\mathbf  R}_{a}(u){\mathbf  -R}_{b}(u^{\prime})\right)  .\label{Hn}%
\end{eqnarray}
Here we used the delta function potential (to make contact with EM), but later
we will show how to generalize to a general correlation. It is useful to
replace the delta function by the equivalent expression:
\begin{equation}
\delta^{(d)}\left(  {\mathbf  R}_{a}(u){\mathbf  -R}_{b}(u^{\prime})\right)
=\int\frac{d{\mathbf  k}}{(2\pi)^{d}}\exp\left(  i{\mathbf  k}\cdot\left(
{\mathbf  R}_{a}(u){\mathbf  -R}_{b}(u^{\prime})\right)  \right)  .
\end{equation}
For a general correlation (see Eq.(\ref{Vf})) we can write
\begin{eqnarray}
&& f\left(  ({\mathbf  R}_{a}(u){\mathbf  -R}_{b}(u^{\prime}))^{2}/d\right)
\nonumber\\
&&= \int d{\mathbf  y\ }f({\mathbf  y}^{2}/d)\int\frac{d{\mathbf  k}}{(2\pi)^{d}}%
\exp(-i{\mathbf  k\cdot y})\ \exp\left(  i{\mathbf  k\cdot}\left(  {\mathbf  R}%
_{a}(u){\mathbf  -R}_{b}(u^{\prime})\right)  \right)  .\label{gf}%
\end{eqnarray}

In order to proceed we use a quadratic variational Hamiltonian to be the best
approximation to $H_{n}$. This is given by
\begin{eqnarray}
h_{n}  & =\int_{0}^{L}du\ \sum_{a=1}^{n}\left[  \frac M2\left(  \frac
{\partial{\mathbf  R}_{a}(u)}{\partial u}\right)  ^{2}+\frac\mu2{\mathbf  R}%
_{a}^{2}(u)\right] \nonumber\\
& -\frac12\int_{0}^{L}du\int_{0}^{L}du^{\prime}\ \sum_{ab}q_{ab}(u-u^{\prime
}){\mathbf  R}_{a}(u)\cdot{\mathbf  R}_{b}(u^{\prime}),\label{hn}%
\end{eqnarray}
where $q_{ab}(u)$ are $n\times n$ variational functions to be determined, with
$n\rightarrow0$ at the end. The best variational Hamiltonian is determined by
the stationarity of the variational free energy which is given
by\cite{feynman,EM,mp}:
\begin{equation}
n\left\langle F\right\rangle _{R}=\left\langle H_{n}-h_{n}\right\rangle
_{h_{n}}-\frac1\beta\ln\int{\mathcal D}{\mathbf  R}_{1}\cdots{\mathcal D}%
{\mathbf  R}_{n}\exp(-\beta h_{n}).\label{fav}%
\end{equation}
The general equations satisfied by $q_{ab}(u)$ where discussed in Refs.
\cite{yygold,chen,yadin}. We showed that although the diagonal elements
$q_{aa}(u) $ must depend on the arc-length variable $u$, the off diagonal
elements $q_{a\neq b}$ which are spin-glass like order parameters can to be
chosen to be $u$- independent; in other words there is a consistent solution
of the variational equations with these properties. (The existence of a
time-persistent part to the off diagonal elements of $q_{ab}$ is well known in
the investigation of quantum spin glass systems \cite{yglai} and is crucial
for the capture of the correct physics in such systems.) In Refs.
\cite{yygold,yadin} we proceeded to solve the equations approximately for the
case of a non-zero confining harmonic potential characterized by a spring
constant $\mu\neq0$. For a quantum particle at not too low a temperature
(equivalent for moderate values of $L$ in the polymer problem) we obtained a
numerical solution of the equations \cite{yygold} for different types of
correlations of the random potential. In Ref. \cite{yadin} we considered the
limit of large $L$ and finite $\mu$ (in the context of the vortex line
problem), and for $d=2$, under certain approximations obtained an analytical
solution to first order in $g$ (the strength of the disorder). Here we would
like to consider the whole range of disorder for large $L$ and also
investigate the limit $\mu\rightarrow0$. Our goal is also to make contact with
the calculation of EM. Hence we will start with a somewhat simpler approach
with a finite number of variational parameters in lieu of the infinite number
of such parameters introduced in our previous work. As will turn out this is
appropriate for the current problem and allows us to solve everything
analytically without any further approximations.

EM considered only the case of $\mu=0$ and chose
\begin{equation}
q_{ab}(u-u^{\prime})=-\frac{q^{2}M}9\delta_{ab}\delta(u-u^{\prime})\label{qem}%
\end{equation}
where $q$ is a single variational parameter. (In an Appendix they considered a
slightly more general form but it is still proportional to $\delta
(u-u^{\prime})$ ). Here we claim that we need to introduce static
($u$-independent) off-diagonal elements for $q_{ab}$ and also add a static
diagonal part. This will help capture the correct physics of the problem as in
the case of the quantum spin glass systems mentioned above. Thus we chose:
\begin{equation}
q_{ab}(u-u^{\prime})=-\delta_{ab}\left(  (\lambda-\mu)\ \delta(u-u^{\prime
})+(\lambda_{1}-\lambda)\ /\ L\right)  +(1-\delta_{ab})\ s\ /\ L,\label{qyg}%
\end{equation}
and we have three variational parameters $\lambda$, $\lambda_{1}$, and $s$. The
variables $\lambda$, $\lambda_{1}$ represent two values of $\lambda(\omega
\neq0)$ and $\lambda(\omega=0)$ instead of the general function $\lambda
(\omega)$ introduced in Ref. \cite{yygold} (which involves an infinity of
variational parameters). The variable $s$ represents a ``spin glass'' type
variable which loosely speaking is a measure of ``freezing''.

The variational Hamiltonian now becomes
\begin{eqnarray}
h_{n}  & =\int_{0}^{L}du\ \sum_{a=1}^{n}\left[  \frac M2\left(  \frac
{\partial{\mathbf  R}_{a}(u)}{\partial u}\right)  ^{2}+\frac\lambda
2{\mathbf  R}_{a}^{2}(u)\right] \nonumber\\
& +\frac1{2L}\sum_{ab}p_{ab}\int_{0}^{L}du\int_{0}^{L}du^{\prime}%
\ {\mathbf  R}_{a}(u)\cdot{\mathbf  R}_{b}(u^{\prime}),\label{hnm}%
\end{eqnarray}
with
\begin{equation}
p_{ab}=\left(
\begin{array}
[c]{cccc}%
\lambda_{1}-\lambda & -s & \cdots & -s\\
-s & \lambda_{1}-\lambda & \ddots & \vdots\\
\vdots & \ddots & \ddots & -s\\
-s & \cdots & -s & \lambda_{1}-\lambda
\end{array}
\right)  ,\label{pmat}%
\end{equation}
which reduces to the EM variational Hamiltonian if $p_{ab}=0$ (i.e. if
$\lambda_{1}=\lambda$ and $s=0$ ). For now we consider a replica symmetric
parametrization. We will discuss a possible replica symmetry breaking
parametrization later on. Using this parametrization of $h_{n}$ our task is to
calculate the free energy from equation (\ref{fav}). This is achieved by first
writing down the propagator associated with $\beta h_{n}$:
\begin{equation}
G_{ab}(\omega)=\left\{  \beta((M\omega^{2}+\mu){\mathbf  1}-\widetilde
{{\mathbf  q}}(\omega))\right\}  _{ab}^{-1}\label{prop}%
\end{equation}

with
\begin{equation}
\widetilde{q}_{ab}(\omega)=\int_{0}^{L}du\ q_{ab}(u)\exp(-i\omega u).
\end{equation}
For the function $q_{ab}(u)$ given by Eq. (\ref{qyg}) we find:
\begin{equation}
\widetilde{q}_{ab}(\omega)=-\delta_{ab}\left(  (\lambda-\mu)\ +(\lambda
_{1}-\lambda)\ \delta_{\omega,0}\right)  +(1-\delta_{ab})\ s\ \delta
_{\omega,0},
\end{equation}
and thus
\begin{equation}
G_{ab}(\omega)=\beta^{-1}\left\{  (M\omega^{2}+\lambda+(\lambda_{1}%
-\lambda+s)\ \delta_{\omega,0}){\mathbf  1}-\ s\ \delta_{\omega,0})\right\}
_{ab}^{-1}\ ,\label{gprop}%
\end{equation}
which gives after inverting an $n\times n$ matrix and taking the limit
$n\rightarrow0$,
\begin{eqnarray}
\beta G_{ab}(\omega =0)&=&\frac{\lambda_{1}+2s}{(\lambda_{1}+s)^{2}}%
\delta_{ab}+\frac s{(\lambda_{1}+s)^{2}}(1-\delta_{ab}),\\
\beta G_{ab}(\omega \neq 0)&=&\frac1{M\omega^{2}+\lambda}\delta_{ab}\ .
\end{eqnarray}
Since the interval on which $u$ is defined is finite ($0\leq u\leq L$ ), the
``frequencies'' $\omega$ are discrete and satisfy
\begin{equation}
\omega_{m}=\frac{2\pi}Lm,\ \ \ m=0,\pm1,\pm2,\ldots.
\end{equation}

We can now use the fact that
\begin{equation}
\left\langle {\mathbf  R}_{a}(u)\cdot{\mathbf  R}_{b}(u^{\prime})\right\rangle
\equiv d\ g_{ab}(u-u^{\prime})=\frac dL\sum_{\omega}e^{-i\omega(u-u^{\prime}%
)}G_{ab}(\omega),
\end{equation}
to obtain an expression for the correlation function of interest and for the
free energy. For the correlation function we obtain
\begin{eqnarray}
C(\ell)  =\frac1d\left\langle \left\langle ({\mathbf  R}(\ell){\mathbf  -R}%
(0))^{2}\right\rangle \right\rangle _{R}=\frac1{nd}\sum_{a=1}^{n}\left\langle
({\mathbf  R}_{a}(\ell){\mathbf  -R}_{a}(0))^{2}\right\rangle \nonumber\\
\  =\frac2{nL}\sum_{a=1}^{n}\sum_{\omega}G_{aa}(\omega)\left(
1-e^{-i\omega\ell}\right)  =\frac2{\beta L}\sum_{\omega\neq0}\frac
{1-e^{-i\omega\ell}}{M\omega^{2}+\lambda}.
\end{eqnarray}
For the free energy we find from Eq. (\ref{fav})
\begin{eqnarray}
\frac{n\left\langle F\right\rangle }d  &&={\mathrm const}.+\frac12(\mu
-\lambda)\sum_{a=1}^{n}\sum_{\omega}G_{aa}(\omega)-\frac12\sum_{ab}%
p_{ab}G_{ab}(\omega=0)\nonumber\\
&& -\frac1{2\beta}\sum_{\omega}\mathrm{tr}\ \ln\ {\mathbf  G}(\omega
)\nonumber\\
&& -\frac{\beta gL}{2d}\int_{0}^{L}dz\ \int\frac{d{\mathbf  k}}{(2\pi)^{d}}%
\sum_{ab}\nonumber\\
&& \exp\left(  -\frac{{\mathbf  k}^{2}}{2L}\sum_{\omega}\left[  G_{aa}%
(\omega)+G_{bb}(\omega)-2e^{-i\omega z}G_{ab}(\omega)\right]  \right)
.\label{nF}%
\end{eqnarray}
We now use the formula (see e.g. Gradshteyn and Ryzhik \cite{gr}, Eq.
1.445.2)
\begin{equation}
\frac1L\sum_{\omega}\frac{e^{-i\omega z}}{M\omega^{2}+\lambda}=\frac
1{2\sqrt{M\lambda}}\frac{\cosh(\alpha(1-2z/L))}{\sinh(\alpha)},\ \ \ \alpha
=\frac L2\sqrt{\frac\lambda M}\ ,\ \ 0\leq z\leq L
\end{equation}
to calculate the correlation function and the free energy in the limit
$n\rightarrow0$: For the correlation function we obtain
\begin{equation}
C(\ell)=\frac1{\beta\sqrt{M\lambda}}\left(  \coth\frac L2\sqrt{\frac\lambda
M}-\frac{\cosh(L\sqrt{\lambda/M}\ (1-2\ell/L)\ /2)}{\sinh(L\sqrt{\lambda
/M}\ /2)}\right)  ,\label{cllam}%
\end{equation}
and for the free energy
\begin{eqnarray}
\frac{\beta\left\langle F\right\rangle }{Ld}  & ={\mathrm const}.+\frac
{(\mu-\lambda)}{4\sqrt{M\lambda}}\coth\frac L2\sqrt{\frac\lambda M}+\frac
\mu{2L}\left(  \frac{\lambda_{1}+2s}{(\lambda_{1}+s)^{2}}-\frac1\lambda\right)
\nonumber\\
& \ +\frac1L\ln\sinh\frac L2\sqrt{\frac\lambda M}+\frac1{2L}\ln\left(  1+\frac
s{\lambda_{1}}\right)  +\frac1{2L}\ln\frac{\lambda_{1}}\lambda-\frac1{2L}\frac
s{\lambda_{1}+s}\nonumber\\
& \ -\frac{\beta^{2}g}{2d}\int_{0}^{L}dz\ \int\frac{d{\mathbf  k}}{(2\pi)^{d}%
}\left[  \exp(-{\mathbf  k}^{2}a_{1})-\exp(-{\mathbf  k}^{2}a_{2})\right]
,\label{full}%
\end{eqnarray}
with
\begin{eqnarray}
&&a_{1}  =\frac1{2\beta\sqrt{M\lambda}}\left(  \coth\frac L2\sqrt{\frac\lambda
M}-\frac{\cosh(L\sqrt{\lambda/M}\ (1-2z/L)\ /2)}{\sinh(L\sqrt{\lambda/M}%
\ /2)}\right)  ,\label{a1}\\
&&a_{2}  =\frac1{\beta L}\left(  \frac1{\lambda_{1}+s}-\frac1\lambda\right)
+\frac1{2\beta\sqrt{M\lambda}}\coth\frac L2\sqrt{\frac\lambda M}.\label{a2}%
\end{eqnarray}
Some of the details of the calculation are given in the Appendix. The constant
term does not depend on the variational parameters. So far the calculation has
been exact but now we are interested in the large $L$ limit. Before we proceed
it will be instructive to pause to review the calculation of EM who have
chosen ${\mathbf  p}=0$, i.e. $s=0$ and $\lambda_{1}=\lambda$ (recall that in
their notation $\lambda\propto q^{2}$). They also take $\mu=0 $. In that case
the free energy simplifies to give:
\begin{eqnarray}
\frac{\beta\left\langle F\right\rangle }{Ld}  & ={\mathrm const}.-\frac
\lambda{4\sqrt{M\lambda}}\coth\frac L2\sqrt{\frac\lambda M}+\frac1L\ln
\sinh\frac L2\sqrt{\frac\lambda M}\nonumber\\
& \ -\frac{\beta^{2}g}d\int_{0}^{L/2}dz\ \int\frac{d{\mathbf  k}}{(2\pi)^{d}%
}\left[  \exp(-{\mathbf  k}^{2}a_{1})-\exp(-{\mathbf  k}^{2}a_{2})\right]  ,
\end{eqnarray}
with $a_{1}$ still given by Eq. (\ref{a1}) and
\begin{equation}
a_{2}=\frac1{2\beta\sqrt{M\lambda}}\coth\frac L2\sqrt{\frac\lambda
M}.\label{a2em}%
\end{equation}
We also noticed that since $a_{1}$ is symmetric about the point $z=L/2$ we
have limited the $z$-integration up to $L/2$ and multiplied the integral by 2.
We can now take the limit of large $L$ . It is at the point $z$=$L/2$ that the
integrand vanishes for large $L$. In this limit we find (upon dropping the
constant):
\begin{equation}
\frac{\beta\left\langle F\right\rangle }{Ld}=\frac14\sqrt{\frac\lambda
M}-\frac{\beta^{2}g}d\int_{0}^{\infty}dz\ \int\frac{d{\mathbf  k}}{(2\pi)^{d}%
}\left[  \exp(-{\mathbf  k}^{2}a_{1})-\exp(-{\mathbf  k}^{2}a_{2})\right]
,\nonumber
\end{equation}
with
\begin{equation}
a_{1}=\frac1{2\beta\sqrt{M\lambda}}\left(  1-\exp\left(  -z\sqrt{\lambda
/M}\right)  \right)  ,\ \ \ \ a_{2}=\frac1{2\beta\sqrt{M\lambda}}.
\end{equation}
Notice that the factor of 2 in front of the integral due to the aforementioned
symmetry was missed in Ref. \cite{EM}. This is of no importance since it just
renormalizes the strength of the disorder. The integral over ${\mathbf  k}$ can
now be done to yield:
\begin{equation}
\frac{\beta\left\langle F\right\rangle }{Ld}=\frac14\sqrt{\frac\lambda
M}-\frac{\beta^{2}g}d\left(  \frac{\beta\sqrt{M\lambda}}{2\pi}\right)
^{d/2}\int_{0}^{\infty}dz\left[  \frac1{\left(  1-\exp\left(  -z\sqrt
{\lambda/M}\right)  \right)  ^{d/2}}-1\right]  .
\end{equation}
At this point we realize that the $z$-integral is infrared divergent for any
dimension $d\geq2$. We can trace this back to the short distance singularity
of the Dirac delta function correlation. We thus replace the delta function by
a regularized form:
\begin{equation}
\delta^{(d)}({\mathbf  R})\rightarrow\frac1{(\pi d\xi^{2})^{d/2}}\exp\left(
-\frac{{\mathbf  R}^{2}}{d\xi^{2}}\right)  ,\label{regdel}%
\end{equation}
where $\xi$ is small (we can think of it as the intrinsic diameter of the
polymer thread). Using the representation given by Eq. (\ref{gf}) for the
right hand side and carrying out the \textbf{y}-integration yields:
\begin{eqnarray}
\frac{\beta\left\langle F\right\rangle }{Ld}  &=&\frac14\sqrt{\frac\lambda
M}\nonumber\\
\ &-&\frac{\beta^{2}g}d\int_{0}^{\infty}dz\ \int\frac{d{\mathbf  k}}{(2\pi)^{d}%
}\exp\left(  -\frac d4\xi^{2}{\mathbf  k}^{2}\right)  \nonumber \\
&\times&\left[  \exp
(-{\mathbf  k}^{2}a_{1})-\exp(-{\mathbf  k}^{2}a_{2})\right]  ,\label{fe1}%
\end{eqnarray}
and the integrals are now properly regularized to yield a finite expression.
To find the optimum variational parameter $\lambda$, we take the derivative of
the above expression with respect to $\sqrt{\lambda}:$%
\begin{eqnarray}
1  & =&\frac{2\beta g}{d\lambda}\sqrt{\frac M\lambda}\int_{0}^{\infty}%
d\tau\ \int\frac{d{\mathbf  k}}{(2\pi)^{d}}{\mathbf  k}^{2}\exp\left(  -\frac
d4\xi^{2}{\mathbf  k}^{2}\right) \nonumber\\
\ &\times&\left[  \left(  1-\exp\left(  -\tau\right)  -\tau\exp\left(
-\tau\right)  \right)  \exp(-{\mathbf  k}^{2}a_{1})-\exp(-{\mathbf  k}^{2}%
a_{2})\right]  ,
\end{eqnarray}
and the $z$-variable has been rescaled by $z\rightarrow\tau\sqrt{M/\lambda}$.
The \textbf{k}-integration can now be done to yield:
\begin{equation}
1=\frac{2g\beta^{d/2+2}M^{d/4+1}}{(2\pi)^{d/2}}\lambda^{\frac{d-4}4}\int
_{0}^{\infty}d\tau\left\{  \frac{1-e^{-\tau}-\tau e^{-\tau}}{\left(
1-e^{-\tau}+\Delta\right)  ^{d/2+1}}-\frac1{(1+\Delta)^{d/2+1}}\right\}
,\label{eqlam}%
\end{equation}
with $\Delta=\xi^{2}d\beta\sqrt{M\lambda}/2$. At this point we see that the
integral is finite for $d<4$ even in the limit $\Delta\rightarrow0$ (which
follows from $\xi\rightarrow0$). Let us denote the integral in this limit by
$I_{d}$:
\begin{equation}
I_{d}=\int_{0}^{\infty}d\tau\left\{  \frac1{\left(  1-e^{-\tau}\right)
^{d/2}}-\frac{\tau e^{-\tau}}{\left(  1-e^{-\tau}\right)  ^{d/2+1}}-1\right\}
,
\end{equation}
so, Eq. (\ref{eqlam}) becomes
\begin{equation}
1=\frac{2g\beta^{d/2+2}M^{d/4+1}}{(2\pi)^{d/2}}I_{d}\ \lambda^{\frac{d-4}%
4}\label{plam}%
\end{equation}

Unfortunately EM did not realize that this integral is \textit{negative} for
$d=3$ (and also $d=2$). The indefinite integral can be carried out
analytically (e.g. using Mathematica) in $d=3$ to give%

\begin{eqnarray}
I_{3}(\tau)  & =-\tau+\frac23\sqrt{1-e^{-\tau}}(-\frac1{1-e^{-\tau}}+\frac
\tau{(1-e^{-\tau})^{2}})+\frac13\ln\frac{1+\sqrt{1-e^{-\tau}}}{1-\sqrt
{1-e^{-\tau}}},\nonumber\\
I_{3}  & =\lim_{\tau\rightarrow\infty}I(\tau)-\lim_{\tau\rightarrow0}%
I(\tau)=-\frac23(1-\ln2)\approx-0.20457.
\end{eqnarray}
(In $d=2$, one has $I_{2}(\tau)=\tau/(e^{\tau}-1)$, and $I_{2}=-1$). From here
it follows that Eq.(\ref{plam}) \textit{has no solution for} $\lambda$\ ! This
is a very important observation. Notice that all fractional powers of
$\lambda$ are always to be taken as positive. For example in the integral
\begin{eqnarray}
\int\frac{d{\mathbf  k}}{(2\pi)^{d}}{\mathbf  k}^{2}\exp(-{\mathbf  k}^{2}a_{2})  &
=\int\frac{d{\mathbf  k}}{(2\pi)^{d}}{\mathbf  k}^{2}\exp\left(  -{\mathbf  k}%
^{2}\frac1{2\beta\sqrt{M\lambda}}\right) \nonumber\\
& =\frac d{2^{d+1}\pi^{d/2}}(2\beta\sqrt{M\lambda})^{d/2+1},
\end{eqnarray}
which is part of the result derived above, $\sqrt{\lambda}$ in the integrand
is positive, and so must be the result of the integration since the integrand
is positive definite. There is no way to argue that $\lambda^{1/4}$ can be
taken as the negative square root of $\sqrt{\lambda}$. To elucidate further
the fact that there is no value of $\lambda$ which extremize the variational
free energy we return to Eq. (\ref{fe1}) of the free energy and carry out the
${\mathbf  k}$-integration to find for $d=3$%
\begin{eqnarray}
\frac{\beta\left\langle F\right\rangle }{3L}  & =&\frac14\sqrt{\frac\lambda
M}-\frac{g\beta^{7/2}M^{5/4}\lambda^{1/4}}{3(2\pi)^{3/2}}\nonumber\\
& \times&\int_{0}^{\infty}d\tau\left(  \frac1{(1+\Delta-\exp(-\tau))^{3/2}%
}-\frac1{(1+\Delta)^{3/2}}\right)  ,\label{fe2}%
\end{eqnarray}
with $\Delta=\xi^{2}d\beta\sqrt{M\lambda}/2$. Again the integral can be done
analytically (Mathematica) and we find
\begin{equation}
\frac{\beta\left\langle F\right\rangle }{3L}=\frac14\sqrt{\frac\lambda
M}-\frac{g\beta^{7/2}M^{5/4}\lambda^{1/4}}{3(2\pi)^{3/2}}\left(
-2(1-\ln2)+\frac2{\sqrt{\Delta}}+O(\sqrt{\Delta})\right)  ,
\end{equation}
substituting for $\Delta$ one obtains
\begin{equation}
\frac{\beta\left\langle F\right\rangle }{3L}=-\frac{g\beta^{3}M}{(3\pi
)^{3/2}\xi}+\frac14\sqrt{\frac\lambda M}+(1-\ln2)\frac{2g\beta^{7/2}%
M^{5/4}\lambda^{1/4}}{3(2\pi)^{3/2}}+O(\xi).
\end{equation}
We see that the divergent term (as $\xi\rightarrow0$) is independent of
$\lambda$. We also see that the free energy is a monotonically increasing
function of $\lambda$ and thus has no extrema as a function of it. Derivative
of the last expression with respect to $\sqrt{\lambda}$ agrees with our
previous result. Thus we see that the one parameter variational Hamiltonian
does not yield a meaningful result.

Let us now return to the more general expression for the variational free
energy given in Eq. (\ref{full}). Before we consider the large $L$ limit we
can draw a general conclusion. Let us calculate the derivative of the fee
energy with respect to $\lambda_{1}$ and $s$:
\begin{eqnarray}
&&\frac{-\mu\lambda_{1}-3\mu s+3s\lambda_{1}+2s^{2}+\lambda_{1}^{2}}{2L\left(
\lambda_{1}+s\right)  ^{3}} \nonumber\\
&&=\left(  \frac{\partial a_{2}}{\partial\lambda_{1}}\right)  \frac{\beta
^{2}g}{2d}\int_{0}^{L}dz\ \int\frac{d{\mathbf  k}}{(2\pi)^{d}}{\mathbf  k}^{2}%
\exp(-{\mathbf  k}^{2}a_{2}),\label{sdif}\\
&&\frac{s\ (2\mu-\lambda_{1}-s)}{2L\left(  \lambda_{1}+s\right)  ^{3}}  
=\left(  \frac{\partial a_{2}}{\partial s}\right)  \frac{\beta^{2}g}{2d}%
\int_{0}^{L}dz\ \int\frac{d{\mathbf  k}}{(2\pi)^{d}}{\mathbf  k}^{2}%
\exp(-{\mathbf  k}^{2}a_{2}).\label{ldif}%
\end{eqnarray}
Since
\begin{equation}
\left(  \frac{\partial a_{2}}{\partial\lambda_{1}}\right)  +\left(
\frac{\partial a_{2}}{\partial s}\right)  =0,
\end{equation}
we find that upon adding the two equations we get
\begin{equation}
\frac{\lambda_{1}-\mu+s}{2L\left(  \lambda_{1}+s\right)  ^{2}}=0,
\end{equation}
which implies
\begin{equation}
\lambda_{1}+s=\mu.
\end{equation}
This is an important general result. Substituting this result in
Eq.(\ref{sdif}) we find
\begin{equation}
s=\frac{\beta g}dL\int\frac{d{\mathbf  k}}{(2\pi)^{d}}{\mathbf  k}^{2}%
\exp(-{\mathbf  k}^{2}a_{2})=\frac{2\pi\beta gL}{(4\pi a_{2})^{d/2+1}},
\end{equation}
with
\begin{equation}
a_{2}=\frac1{\beta L}\left(  \frac1\mu-\frac1\lambda\right)  +\frac
1{2\beta\sqrt{M\lambda}}\coth\frac L2\sqrt{\frac\lambda M},
\end{equation}
which gives a relation between $s$ and $\lambda$. However as we will see in a
moment, only the combination $\lambda_{1}+s=\mu$, enters in the equation for
$\lambda$.

Returning to Eq. (\ref{full}) we see that upon taking the derivative with
respect to $\sqrt{\lambda}$ the only dependence on $s$ and $\lambda_{1}$ is
through the combination $\lambda_{1}+s$ in $a_{2}$. It is simpler to take the
limit of large $L$ and to write up the resulting equation for $\lambda$ up to
exponentially small terms in $L:$%
\begin{eqnarray}
\lambda-\mu & =\frac{4(\lambda-\mu)}L\sqrt{\frac M\lambda}+\frac{2\beta
g}d\sqrt{\frac M\lambda}\int_{0}^{\frac L2\sqrt{\frac\lambda M}}d\tau\int
\frac{d{\mathbf  k}}{(2\pi)^{d}}{\mathbf  k}^{2}\nonumber\\
& \ \times\left(  \exp\left(  -{\mathbf  k}^{2}a_{1}\right)  \left(
1-\exp\left(  -\tau\right)  -\tau\exp\left(  -\tau\right)  \right)  \right.
\nonumber\\
& \ \left.  -\exp\left(  -{\mathbf  k}^{2}a_{2}\right)  \left(  1-4\sqrt
{M/\lambda}/L\right)  \right)  ,
\end{eqnarray}
with
\begin{equation}
a_{1}=\frac{1-e^{-\tau}}{2\beta\sqrt{M\lambda}},\ \ \ a_{2}=\frac1{2\beta
\sqrt{M\lambda}}+\frac1{\beta L\mu}-\frac1{\beta L\lambda}.
\end{equation}
If $\mu$ is finite, one can proceed with expanding $\exp\left(  -{\mathbf  k}%
^{2}a_{2}\right)  $ in powers of $1/L$ as will be done later. However if one
attempts to take the limit $\mu\rightarrow0$ we see immediately a potential
problem because of the term $1/(\beta L\mu)$ in $a_{2}$. If we carry out the
${\mathbf  k}$-integration we find
\begin{eqnarray}
\lambda-\mu &=&\frac{2g\beta^{d/2+2}M^{d/4+1}}{(2\pi)^{d/2}}\lambda^{\frac
d4}\nonumber\\
 \ \times&\int_{0}^{\frac L2\sqrt{\frac\lambda M}}&d\tau\left(  \frac
{1-e^{-\tau}-\tau e^{-\tau}}{\left(  1-e^{\tau}\right)  ^{d/2+1}}%
-\frac1{\left(  1+\frac2L\sqrt{\frac M\lambda}\frac{\lambda-\mu}\mu\right)
^{d/2+1}}\right)  ,
\end{eqnarray}
where we have omitted subleading terms in $1/L$. As $\mu\rightarrow0$ for
fixed large $L$, the last term in the integral vanishes (as $\lambda-\mu$
remains finite for $g\neq0$). The integral over $\tau$ no longer converges for
large $L$, but is rather proportional to $L$. To leading order we get (by
subtracting and adding 1 to the integrand)
\begin{equation}
\lambda=\frac{g\beta^{d/2+2}M^{d/4+1/2}L}{(2\pi)^{d/2}}\lambda^{\frac{d+2}%
4},\label{lammu0}%
\end{equation}
which gives
\begin{equation}
\lambda=\left(  \frac{g\beta^{d/2+2}M^{d/4+1/2}L}{(2\pi)^{d/2}}\right)
^{\frac4{2-d}}.
\end{equation}
We see that the borderline dimension appears to be $d=2$. Indeed from Eq.
(\ref{cllam}) it follows that for large $L$%
\begin{equation}
C(\ell)\approx\frac1{\beta\sqrt{M\lambda}}\left(  1-\exp(-\ell\sqrt{\lambda
/M})\right)  ,
\end{equation}
and thus the radius of gyration satisfies
\begin{equation}
R\sim\lambda^{-1/4}\sim(gL)^{-\frac1{2-d}},
\end{equation}
which agrees perfectly with equation (\ref{Rl2}) for $d<2$. To see what
happens for $d>2$ we can easily show that in the limit $\mu\rightarrow0$ the
free energy becomes of the form
\begin{equation}
\frac{\beta\left\langle F\right\rangle }{Ld}=\frac14\sqrt{\frac\lambda
M}-\frac{\beta^{2}g}d\left(  \frac{\beta\sqrt{M\lambda}}{2\pi}\right)
^{d/2}\sqrt{\frac M\lambda}\int_{0}^{\frac L2\sqrt{\frac\lambda M}}d\tau
\frac1{\left(  1+\Delta-\exp\left(  -\tau\right)  \right)  ^{d/2}},
\end{equation}
where again we regularized with $\Delta=\xi^{2}d\beta\sqrt{M\lambda}/2$. This
gives
\begin{equation}
\frac{\beta\left\langle F\right\rangle }{Ld}={\mathrm const}.+\frac
14\sqrt{\frac\lambda M}-\frac{\beta^{2}g}{2d}\left(  \frac{\beta\sqrt
{M\lambda}}{2\pi}\right)  ^{d/2}L,
\end{equation}
and when using $\lambda\sim d^{2}\beta^{-2}M^{-1}R^{-4}$ we obtain
\begin{equation}
\beta\left\langle F\right\rangle ={\mathrm const}.\times L+\frac{d^{2}}{4\beta
M}\frac L{R^{2}}-\frac{d^{d/2}\beta^{2}g}{2(2\pi)^{d/2}}\frac{L^{2}}{R^{d}},
\end{equation}
which coincides with Eq. (\ref{fran}) and shows that for $d>2$, $F\rightarrow
-\infty$ as $R\rightarrow0$ and there is always collapse. Thus we see that in
the limit of $\mu\rightarrow0$ we recover the annealed result from the replica
calculation as expected.

If on the other hand $\mu$ is finite, we can expand $\exp\left(
-{\mathbf  k}^{2}a_{2}\right)  $ in powers of $1/L$ and we find to leading order
in $L$:
\begin{eqnarray}
&&\lambda-\mu =\frac{2\beta g}{d}\sqrt{\frac{M}{\lambda}}\int_{0}^{\infty
}d\tau\int\frac{d{\mathbf  k}}{(2\pi)^{d}}{\mathbf  k}^{2}\nonumber\\
&&\times\left(  \exp\left(  -{\mathbf  k}^{2}\frac{1-e^{-\tau}}%
{2\beta\sqrt{M\lambda}}\right)  \left(  1-\exp\left(  -\tau\right)  -\tau
\exp\left(  -\tau\right)  \right)  -\exp\left(  -\frac{{\mathbf  k}^{2}}%
{2\beta\sqrt{M\lambda}}\right)  \right) \nonumber\\
&& +\frac{4\beta g}{d}\sqrt{\frac{M}{\lambda}}\int\frac{d{\mathbf  k}%
}{(2\pi)^{d}}{\mathbf  k}^{2}\exp\left(  -\frac{{\mathbf  k}^{2}}{2\beta
\sqrt{M\lambda}}\right) \nonumber\\
&& +\frac{g}{d}\left(  \frac{1}{\mu}-\frac{1}{\lambda}\right)
\int\frac{d{\mathbf  k}}{(2\pi)^{d}}{\mathbf  k}^{4}\exp\left(  -\frac
{{\mathbf  k}^{2}}{2\beta\sqrt{M\lambda}}\right)  .\label{lamu}%
\end{eqnarray}
The last two terms are also O(1) although they originated from seemingly $1/L
$ terms, since we obtain a factor of $L$ from the range of integration over
$\tau$. Evaluating the integrals we find
\begin{equation}
\lambda-\mu=\frac{2g\beta^{d/2+2}M^{d/4+1}}{(2\pi)^{d/2}}(I_{d}+2+\frac
{d+2}{2}\frac{\lambda-\mu}{\mu})\ \lambda^{\frac{d}{4}}.\label{lameq}%
\end{equation}
For small $g$ we can solve this equation in powers of $g$. Defining a
dimensionless constant
\begin{equation}
\widetilde{g}=\frac{g(\beta^{2}M)^{(d+4)/4}\mu^{(d-4)/4}}{(2\pi)^{d/2}%
},\label{gtilde}%
\end{equation}
we cast the Eq. (\ref{lameq}) in the form
\begin{equation}
h=1+\widetilde{g}2(I_{d}+\frac{2-d}{2}+\frac{2+d}{2}h)h^{d/4},\label{lamovmu}%
\end{equation}
with $h\equiv\lambda/\mu$. To second order in $\widetilde{g}$ we find:
\begin{equation}
\lambda/\mu=1+2(I_{d}+2)\widetilde{g}+(I_{d}+2)\left(  d(I_{d}%
+2)+2(d+2)\right)  \widetilde{g}^{2}+\ldots.
\end{equation}
Thus as $g$ increases from 0, $\lambda$ is an increasing function of $g$
starting from an initial value of $\mu$. However a numerical solution of
equation (\ref{lameq}) (for $d=3$) reveals that the solution becomes ill
behaved as $\lambda$ becomes of magnitude $\sim2\mu$. This happens for
$\widetilde{g}\sim1/(2^{7/4}(I_{3}+2))$. The reason for this is as will become
evident in the next section is that the replica symmetric solution becomes
invalid at this point and has to be replaced by a replica symmetry breaking
solution. This will become clear in the next section where we will find the
correct solution for larger values of $\widetilde{g}$. It is also clear from
Eq. (\ref{gtilde}) that for fixed $g$ as $\mu\rightarrow0$, $\widetilde{g}$
becomes large and we will be in the region when RSB is to be used. Thus the
range of applicability of the replica symmetric solution is minimal for
a small value of $\mu$.

The rest of the section can be skipped on first reading of the paper and the
interested reader might continue directly to the next section discussing the
RSB solution. For completeness we display here the form Eq.(\ref{lamu}) takes
for a general correlation of the disorder defined in Eq.(\ref{Vf}). We can use
the representation given in Eq. (\ref{gf}) to obtain:
\begin{eqnarray}
&&\lambda-\mu =-4\beta g\sqrt{\frac{M}{\lambda}}\int_{0}^{\infty}%
d\tau\nonumber\\
&& \times\left(  \widehat{f}\ ^{\prime}\left(  \frac{1-e^{-\tau}}{\beta
\sqrt{M\lambda}}\right)  \left(  1-\exp\left(  -\tau\right)  -\tau\exp\left(
-\tau\right)  \right)  -\widehat{f}\ ^{\prime}\left(  \frac{1}{\beta
\sqrt{M\lambda}}\right)  \right) \nonumber\\
&& -8\beta g\sqrt{\frac{M}{\lambda}}\widehat{f}\ ^{\prime}\left(  \frac
{1}{\beta\sqrt{M\lambda}}\right)  +4g\left(  \frac{1}{\mu}-\frac{1}{\lambda
}\right)  \widehat{f}\ ^{\prime\prime}\left(  \frac{1}{\beta\sqrt{M\lambda}%
}\right)  ,
\end{eqnarray}
where we defined
\begin{eqnarray}
\widehat{f}(a)  & \equiv\int d{\mathbf  y\ }f({\mathbf  y}^{2}/d)\int
\frac{d{\mathbf  k}}{(2\pi)^{d}}\exp(-i{\mathbf  k\cdot y})\exp\left(
-\frac{a{\mathbf  k}^{2}}{2}\right) \nonumber\\
\  & =\frac{1}{\Gamma(d/2)}\int_{0}^{\infty}dx\ x^{d/2-1}e^{-x}f\ \left(
\frac{2xa}{d}\right)  ,\label{fhat}%
\end{eqnarray}
and the primes stand for derivatives of $\widehat{f}$, which can be obtained
from the first line of Eq. (\ref{fhat}) by taking the derivative with respect
to $a$ under the integral sign.

At this point we would like to discuss the more complete variational scheme
that we used in Refs. \cite{yygold,yadin} and show that all our conclusions
concerning the limit $\mu\rightarrow0$ follows from that scheme as well. Since
the notation there was different we will translate the equations to the
present notation but we will not rederive them here. What we did there was to
consider a variational scheme in which we allowed the variable $\lambda$ to
depend on $\omega$ and we extremized the free energy with respect to each
variable $\lambda(\omega)$. The propagator $G(\omega)$ defined in
Eq.(\ref{gprop}) now becomes
\begin{equation}
G_{ab}(\omega)=\beta^{-1}\left\{  (M\omega^{2}+\lambda(\omega)+(\lambda
_{1}-\lambda(0)+s)\ \delta_{\omega,0}){\mathbf  1}-\ s\ \delta_{\omega
,0})\right\}  _{ab}^{-1}\ .
\end{equation}
We have found that the relation $\lambda_{1}+s=\mu$ still holds and $s$ and
$\lambda(\omega)$ satisfy the equations
\begin{eqnarray}
&& s=-2\beta gL\widehat{f}\ ^{\prime}(2a_{2}),\\
&& \lambda(\omega)-\mu  =-s-2\beta g\int_{0}^{L}dz\ (1-e^{i\omega z})\widehat
{f}\ ^{\prime}(2a_{1}(z)),\ \ \ \omega\neq0,\label{lamom}%
\end{eqnarray}
with
\begin{eqnarray}
a_{1}(z)  =\frac{1}{\beta L}\sum_{\omega\neq0}\frac{1-e^{-i\omega z}%
}{M\omega^{2}+\lambda(\omega)},\\
a_{2}   =\frac{1}{\beta\mu L}+\frac{1}{\beta L}\sum_{\omega\neq0}\frac
{1}{M\omega^{2}+\lambda(\omega)}.
\end{eqnarray}
For a regularized delta function correlation we have
\begin{equation}
\widehat{f}\ ^{\prime}(a)=-\frac{1}{2(2\pi)^{d/2}}\frac{1}{(d\xi
^{2}/2+a)^{d/2+1}}.
\end{equation}
In the limit $\mu\rightarrow0$, we observe that $s\rightarrow0$, and there is
no longer a cancelation of the contributions linear in $L$ between the two
terms on the right hand side of Eq.(\ref{lamom}). Instead we get
\begin{equation}
\lambda(\omega)=-2\beta gL\widehat{f}\ ^{\prime}\left(  \frac{2}{\beta L}%
\sum_{\omega\neq0}\frac{1}{M\omega^{2}+\lambda(\omega)}\right)  ,
\end{equation}
which yields an $\omega$- independent solution that for the delta correlation
becomes
\begin{equation}
\lambda=\frac{\beta gL}{(2\pi)^{d/2}}(\beta\sqrt{M\lambda})^{d/2+1}.
\end{equation}
This result exactly coincides with Eq.(\ref{lammu0}) derived previously.

\section{Replica symmetry breaking}

In the previous section we have seen that the replica symmetric solution
becomes invalid for fixed amount of disorder and small harmonic constant $\mu
$. In this section we show the emergence of a different solution of the
variational equation which is more adequate for our problem. But in order to
take advantage of such a solution we must use a more general variational
scheme. Returning to Eqs. (\ref{qyg})-(\ref{pmat}), we have extended the
parametrization of the matrix $p_{ab}$ in (\ref{pmat}) to allow for one-step
RSB by having two off-diagonal parameters $s_{0}$ ($x<x_{c}$) and $s_{1}%
$($x>x_{c}$) together with a breaking point $x_{c}\ (0\leq x_{c}\leq1)$ . Here
$x$ is Parisi's replica index. For details of Parisi's RSB scheme see reviews
of spin glass theory \cite{parisi,mpv,by}. Thus our variational scheme
includes now 5 parameters. A one step breaking is sufficient for the case of
short range correlations of the random potential \cite{yygold,engel}.

We were able to calculate analytically the free energy with the new
parameters. Here we display the final result, the details given in the
Appendix :
\begin{eqnarray}
&&\frac{\beta\left\langle F\right\rangle }{Ld}  =\frac{(\mu-\lambda)}%
{4\sqrt{M\lambda}}\coth\frac{L}{2}\sqrt{\frac{\lambda}{M}}+\frac{\mu}%
{2L}\left(  \frac{1}{x_{c}(\lambda_{1}+s_{1}-\Sigma)}+\left(  1-\frac{1}%
{x_{c}}\right)  \frac{1}{\lambda_{1}+s_{1}}\right. \nonumber\\
&& \left.  +\frac{s_{0}}{(\lambda_{1}+s_{1}-\Sigma)^{2}}-\frac{1}{\lambda
}\right)  +\frac{1}{L}\ln\sinh\frac{L}{2}\sqrt{\frac{\lambda}{M}}+\frac{1}%
{2L}\ln\left(  1+\frac{s_{1}-\Sigma}{\lambda_{1}}\right) \nonumber\\
&& \ -\frac{1}{2L}\left(  1-\frac{1}{x_{c}}\right)  \ln\left(  1-\frac{\Sigma
}{\lambda_{1}+s_{1}}\right)  +\frac{1}{2L}\ln\frac{\lambda_{1}}{\lambda}%
-\frac{1}{2L}\frac{s_{0}}{\lambda_{1}+s_{1}-\Sigma}\nonumber\\
&& \ \ -\frac{\beta^{2}g}{2d}\int_{0}^{L}dz\ \int\frac{d{\mathbf  k}}{(2\pi)^{d}%
}\left[  \exp(-{\mathbf  k}^{2}a_{1})-x_{c}\exp(-{\mathbf  k}^{2}a_{2l}%
)-(1-x_{c})\exp(-{\mathbf  k}^{2}a_{2b})\right]  ,\label{fersb}%
\end{eqnarray}
We introduced the notation
\begin{equation}
\Sigma=x_{c}(s_{1-}s_{0}),
\end{equation}
the variable $a_{1}$ is still given by Eq. (\ref{a1}) and we defined
\begin{eqnarray}
&&a_{2l}   =\frac{1}{\beta L}\left(  \frac{1}{x_{c}}\frac{1}{\lambda_{1}%
+s_{1}-\Sigma}+\left(  1-\frac{1}{x_{c}}\right)  \frac{1}{\lambda_{1}+s_{1}%
}-\frac{1}{\lambda}\right)  +\frac{1}{2\beta\sqrt{M\lambda}}\coth\frac{L}%
{2}\sqrt{\frac{\lambda}{M}},\label{a2lf}\\
&&a_{2b}   =\frac{1}{\beta L}\left(  \frac{1}{\lambda_{1}+s_{1}}-\frac
{1}{\lambda}\right)  +\frac{1}{2\beta\sqrt{M\lambda}}\coth\frac{L}{2}%
\sqrt{\frac{\lambda}{M}}.\label{a2bf}%
\end{eqnarray}
From the free energy we are able to get the following five relations
(everywhere we eliminated $s_{1}$ in favor of $\Sigma$)
\begin{equation}
\lambda_{1}+s_{0}-(1-1/x_{c})\Sigma=\mu,\label{l1eq}%
\end{equation}
which replaces the relation $\lambda_{1}+s=\mu$ established above for the
replica symmetric solution.
\begin{equation}
s_{0}=\frac{\beta Lg}{d}\int\frac{d{\mathbf  k}}{(2\pi)^{d}}{\mathbf  k}^{2}%
\exp(-{\mathbf  k}^{2}a_{2l}),\label{s0eq}%
\end{equation}%
\begin{equation}
\Sigma=\frac{\beta g}{d}Lx_{c}\int\frac{d{\mathbf  k}}{(2\pi)^{d}}{\mathbf  k}%
^{2}\left[  \exp(-{\mathbf  k}^{2}a_{2b})-\exp(-{\mathbf  k}^{2}a_{2l})\right]
,\label{sigeq}%
\end{equation}%
\begin{eqnarray}
\frac{\Sigma}{\mu+\Sigma}-\ln\left(  1+\frac{\Sigma}{\mu}\right) 
=\frac{\beta g}{d}Lx_{c}\frac{\Sigma}{\mu(\mu+\Sigma)}\int\frac{d{\mathbf  k}%
}{(2\pi)^{d}}{\mathbf  k}^{2}\exp(-{\mathbf  k}^{2}a_{2l})\nonumber\\
 \ +\frac{\beta^{2}g}{d}(Lx_{c})^{2}\int\frac{d{\mathbf  k}}{(2\pi)^{d}}\left[
\exp(-{\mathbf  k}^{2}a_{2l})-\exp(-{\mathbf  k}^{2}a_{2b})\right]  ,\label{xeq}%
\end{eqnarray}%
\begin{eqnarray}
&&\lambda-\mu =\frac{2\beta g}{d}\sqrt{\frac{M}{\lambda}}\int_{0}^{\frac{L}%
{2}\sqrt{\frac{\lambda}{M}}}d\tau\int\frac{d{\mathbf  k}}{(2\pi)^{d}}%
{\mathbf  k}^{2}\nonumber\\
&& \times\left(  \exp\left(  -{\mathbf  k}^{2}a_{1}\right)  \left(
1-\exp\left(  -\tau\right)  -\tau\exp\left(  -\tau\right)  \right)  \right.
\nonumber\\
&& \left.  \left(  -x_{c}\exp\left(  -{\mathbf  k}^{2}a_{2l}\right)
-(1-x_{c})\exp\left(  -{\mathbf  k}^{2}a_{2b}\right)  \right)  \left(
1-4\sqrt{M/\lambda}/L\right)  \right)  ,\label{lamdaeq}%
\end{eqnarray}
where we defined
\begin{eqnarray}
&&a_{1}   =\frac{1-e^{-\tau}}{2\beta\sqrt{M\lambda}}\label{a2b}\\
&&a_{2b}  =\frac{1}{2\beta\sqrt{M\lambda}}-\frac{1}{\beta L\lambda}+\frac
{1}{\beta L(\mu+\Sigma)},\\
&&a_{2l}  =\frac{1}{2\beta\sqrt{M\lambda}}-\frac{1}{\beta L\lambda}+\frac
{1}{\beta\mu}\frac{1}{Lx_{c}}+\frac{1}{\beta(\mu+\Sigma)}\left(  \frac{1}%
{L}-\frac{1}{Lx_{c}}\right)  .\label{a2l}%
\end{eqnarray}
We have simplified some expressions assuming large $L$ and dropped a term of
order $1/L$ in Eq. (\ref{lamdaeq}).

If we denote by $y_{c}=Lx_{c}$ we realize that equations (\ref{sigeq}) and
(\ref{xeq}) can be solved for $\Sigma$ and $y_{c}$ of $O(1)$ with respect to
$L$. These equations are similar for those of a classical particle in a random
potential, \cite{engel} except for the variable $\lambda$ which does not
appear there. (One can recover the equations for the classical particle by
taking the limit $M\rightarrow\infty$ with $L$ fixed. One needs to replace
$\beta L$ with $\beta$ for a particle. This limit is not meaningful for a
polymer.) For small $\mu$ we can have an approximate analytical solution:
\begin{eqnarray}
&&\Sigma  =\frac{\sqrt{gd}}{(2\pi)^{d/4}}(\beta\sqrt{M\lambda})^{d/4+1}%
\sqrt{\left|  \ln\mu\right|  },\label{suga}\\
&& y_{c}  \equiv Lx_{c}=\frac{1}{\beta}\sqrt{\frac{d}{g}}(2\pi)^{d/4}%
(\beta\sqrt{M\lambda})^{-d/4}\sqrt{\left|  \ln\mu\right|  },\label{xa}\\
&& s_{0}  ={\mathrm const}.\times g^{(2-d)/4}\beta L(\beta\sqrt{M\lambda
})^{-d(d+2)/8}\mu^{d/2+1}\left|  \ln\mu\right|  ^{(d+2)/4}.\label{s0a}%
\end{eqnarray}
An analysis of the equations (expanding in power series in $\Sigma$) shows
that this solution is valid as long as the condition
\begin{equation}
2\beta\sqrt{M\lambda}\left(  \frac{g(2+d)}{2^{2+d}\pi^{d/2}}\right)
^{2/(4+d)}\mu^{-4/(4+d)}\geq1\label{condrsb}%
\end{equation}
is satisfied. This inequality can also be expressed in the form
\begin{equation}
\widetilde{g}(d+2)\left(  \frac{\lambda}{\mu}\right)  ^{(d+4)/4}%
\geq1,\label{crsb}%
\end{equation}
where $\widetilde{g}$ has been defined in Eq. (\ref{gtilde}). When the
equality holds we have $\Sigma=0$ and $x_{c}=(4+d)\sqrt{M\lambda}/(2\mu L)$.
This can also be verified by using this condition at the equality point in the
above solutions for $\Sigma$ and $x_{c}$ and we see that indeed $\Sigma\sim
O(\mu)$, and $x_{c}\sim\sqrt{M\lambda}/(\mu L)$. Solving the equality
condition given by Eq. (\ref{crsb}) together with Eq. (\ref{lamovmu}) gives
\begin{eqnarray}
&&h =1+\sqrt{\frac{4+2I_{d}}{d+2}}\approx1.85\ \mathrm{for\ }d=3,\\
&&\widetilde{g} \approx\left(  (d+2)h^{(d+4)/4}\right)  ^{-1}\simeq
0.068\ \mathrm{for\ }d=3,
\end{eqnarray}
in agreement with our numerical solution of Eq.(\ref{crsb}) which broke down
for $h\approx2$ for $d=3$. So the point when the replica symmetric solution
has to be replaced by the RSB solution is just below the point that the RS
solution becomes ill behaved.

If on the other hand $\mu$ is small but fixed we can use the solution we have
obtained above in the equation for $\lambda$, in the limit of large $L$. We
obtain
\begin{eqnarray}
&&\lambda-\mu  =\frac{2\beta g}{d}\sqrt{\frac{M}{\lambda}}\int_{0}^{\infty
}d\tau\int\frac{d{\mathbf  k}}{(2\pi)^{d}}{\mathbf  k}^{2}\nonumber\\
&& \times\left(  \exp\left(  -{\mathbf  k}^{2}a_0(1-e^{-\tau})%
\right)  \left(  1-\exp\left(  -\tau\right)  -\tau
\exp\left(  -\tau\right)  \right)  -\exp\left(  -{\mathbf  k}^{2}a_0%
\right)  \right) \nonumber\\
&& -\frac{\beta gy_{c}}{d}\int\frac{d{\mathbf  k}}{(2\pi)^{d}%
}{\mathbf  k}^{2}\left(  \exp\left(  -{\mathbf  k}^{2}\left( %
a_0+\frac{\Sigma}{\beta y_{c}\mu(\mu+\Sigma)}\right)
\right)  -\exp\left(  -{\mathbf  k}^{2}a_0\right)
\right) \nonumber\\
&& +\frac{4\beta g}{d}\sqrt{\frac{M}{\lambda}}\int\frac{d{\mathbf  k}%
}{(2\pi)^{d}}{\mathbf  k}^{2}\exp\left(  -{\mathbf  k}^{2}a_0
\right) \nonumber\\
&& +\frac{g}{d}\left(  \frac{1}{\mu+\Sigma}-\frac{1}{\lambda}\right)  \int
\frac{d{\mathbf  k}}{(2\pi)^{d}}{\mathbf  k}^{4}\exp\left(  -{\mathbf  k}^{2}a_0%
\right) ,
\end{eqnarray}
where we defined
\begin{equation}
a_0\equiv 1/(2\beta\sqrt{M\lambda}).
\end{equation}
We can check that for $\Sigma=0$ it reduces to the replica symmetric equation.
Carrying out the integrals we get
\begin{eqnarray}
&&\lambda-\mu =\frac{2g\beta^{d/2+2}M^{d/4+1}}{(2\pi)^{d/2}}\nonumber\\
&& \times\left(  I_{d}+2+\frac{y_{c}}{2}\sqrt{\frac{\lambda}{M}}\left(
1-\left(  1+\frac{2\Sigma\sqrt{M\lambda}}{y_{c}\mu(\mu+\Sigma)}\right)
^{-d/2-1}\right)  +\frac{d+2}{2}\frac{\lambda-\mu-\Sigma}{\mu+\Sigma}\right)
\ \lambda^{\frac{d}{4}},\label{rsblambda}%
\end{eqnarray}
and we have to substitute for $\Sigma$ and $y_{c}$ (which are functions of
$\lambda$) from Eqs. (\ref{suga}) and (\ref{xa}) respectively. If we now
consider the case of strong disorder we can neglect $\mu$ relative to
$\lambda$ and $\Sigma$ and the above equation simplifies to give
\begin{equation}
\lambda^{1-d/4}=\frac{2g\beta^{d/2+2}M^{d/4+1}}{(2\pi)^{d/2}}\left(
I_{d}+2+\frac{y_{c}}{2}\sqrt{\frac{\lambda}{M}}+\frac{d+2}{2}\left(
\frac{\lambda}{\Sigma}-1\right)  \right)  .
\end{equation}
Substituting for $\Sigma$ and $y_{c}$ we find
\begin{eqnarray}
&&\lambda^{(4-d)/4} =\frac{2g\beta^{d/2+2}M^{d/4+1}}{(2\pi)^{d/2}}\left(
I_{d}+\frac{2-d}{2}\right. \nonumber\\
&& \left.  +\frac{\sqrt{d/4}}{\sqrt{g}}(2\pi)^{\frac{1}{4}d}\beta
^{-(4+d)/4}M^{-(4+d)/8}\lambda^{(4-d)/8}\sqrt{\left|  \ln\mu\right|  }\left(
1+\frac{d+2}{d\left|  \ln\mu\right|  }\right)  \right)  .\label{lamint}%
\end{eqnarray}
Let us seek a solution of the form
\begin{equation}
\lambda=C^{8/(4-d)}\left(  \beta^{2}M\right)  ^{(4+d)/(4-d)}g^{4/(4-d)}.
\end{equation}
Substituting in Eq. (\ref{lamint}) we obtain a quadratic equation for $C$ and
to leading order as $\mu\rightarrow0$ we find:
\begin{equation}
\lambda=\frac{d^{4/(4-d)}}{(2\pi)^{2d/(4-d)}}\left(  \beta^{2}M\right)
^{(4+d)/(4-d)}\left(  g\ \left|  \ln\mu\right|  \right)  ^{4/(4-d)}%
.\label{lamfin}%
\end{equation}
Using this result inside the parenthesis in Eq. (\ref{lamint}) we see that we
get $I_{d}+2+\frac{1}{2}d\left|  \ln\mu\right|  $ . This shows that we were
justified \textit{a posteriori} in neglecting the constant terms. It also
shows that the negative constant $I_{d}$ of EM (see Eq. \ref{plam}) has been
replaced by the term $\frac{1}{2}d\left|  \ln\mu\right|  $ . From this final
result we obtain the radius of gyration
\begin{eqnarray}
R  & \sim(\beta^{2}M\lambda/d^{2})^{-1/4}=\left(  \frac{d^{(d-2)/2}}%
{(2\pi)^{d/2}}\beta^{4}M^{2}g\ \left|  \ln\mu\right|  \right)  ^{-1/(4-d)}%
\nonumber\\
\  & \sim\left(  \frac{4d^{d/2}}{(2\pi)^{d/2}b^{4}}\beta^{2}g\ \ln V\right)
^{-1/(4-d)}.\label{rfin}%
\end{eqnarray}

This is the main result of the paper. It recovers the EM result but with their
constant $I_{d}$ being replaced by $2\ln V$ as has been argued by Cates and
Ball \cite{cates}. Note that we have replaced $M$ in favor of the bond step
$b$.

Substituting the result (\ref{lamfin}) obtained for $\lambda$ in Eqs.
(\ref{suga} and (\ref{xa}) we find
\begin{eqnarray}
\Sigma & =\left(  \frac{d}{(2\pi)^{d/2}}g\beta^{(d+4)/2}M^{(d+4)/4}\left|
\ln\mu\right|  \right)  ^{4/(4-d)},\\
y_{c}  & =Lx_{c}=\left(  \frac{d^{d-2}}{(2\pi)^{d}}g^{2}\beta^{d+4}%
M^{d}\left|  \ln\mu\right|  ^{d-2}\right)  ^{-1/(4-d)}.\label{ycf}%
\end{eqnarray}
The second equation is important since $x_{c}$ can not exceed $1$ (Parisi's
variable $x$ must satisfy $0\leq x\leq1$ \cite{parisi}) . For $2<d<4$ we see
that $x_{c}$ actually decreases when $\mu$ becomes small so there is no
problem. Also for $d=2$ there is no problem since for large enough $L$,
$x_{c}$ is also within range. On the other hand when $d<2$, $x_{c}$ increases
when $\mu$ becomes small (or equivalently $V$ becomes large) and eventually
will exceed 1. For example for $d=1$ we see that this happens for $\left|
\ln\mu\right|  \sim g^{2}L^{3}$, which corresponds to an extremely large
volume $V^{\prime}$ $\sim\exp(g^{2}L^{3})$ when $L$ is large. For
$V>V^{\prime}$ we revert to the annealed result, which for $d<2$ predict
$R\sim(Lg)^{1/(d-2)}$ as was shown in the last section. In the large $L$ limit
this again leads to a fully collapsed polymer.

We have also verified that to leading order the free energy is given by Eq.
(\ref{FQ}) (there is a subleading term of the form $Lg/R^{d-2}$ that can be
neglected). It is interesting that the condition $x_{c}<1$ that we have
applied above has a physical significance \cite{cates}. The attractive term in
the free energy is (see Eq. (\ref{FQ})) of the form $-L\sqrt{g\ln V/R^{d}}$.
This represents (up to a sign) the binding energy of the chain. In order that
the polymer will be confined to a small single region of size $R$ as given
above in Eq. (\ref{rfin}), the binding energy should not exceed the
translational entropy $\sim\ln V$. The condition
\begin{equation}
\ln V<L\sqrt{g\ln V/R^{d}}\label{ineq}%
\end{equation}
is equivalent (up to some irrelevant constants) to the condition $x_{c}<1$ as
can be verified by using the result (\ref{rfin}) in Eq. (\ref{ineq}).

\section{Conclusions}

We have considered the problem of a polymer ( a Gaussian chain) in a quenched
disordered medium. The problem maps also to a quantum particle in a random
potential, and in the presence of an additional confining harmonic force (of
spring constant $\mu$ ) it maps also to the problem of a flux line in a cage
potential and random columnar disorder. We carried out a replica calculation
in the presence of a confining harmonic force, and succeeded to ``improve''
the previous results of EM \cite{EM}, in the sense that the (unphysical)
constant is replaced by $\ln V$ in the equation for the
variational parameter $\lambda$ and hence also in the dependence of the radius
of gyration on the strength of the disorder. Of course our calculation does
not diminish the accomplishments of EM who pioneered the use of the
variational method in the context of the replica calculation and for the first
time obtained the correct scaling exponent for the dependence of the radius of
gyration on the disorder for finite systems. In the infinite volume limit the
chain collapses since it can find a very deep potential minimum somewhere
which can accommodate it. For $2\leq d\leq4 $ the chain is \textit{localized}
in the sense \cite{cates} that even in the large $V$ limit two long chains
introduced into the system will find the same small neighborhood to occupy
(with a probability approaching 1 for large $L$). This is a consequence of the
off diagonal spin-glass order parameter we introduced that measures overlap
between different replicas. It is comforting to find out that the replica
calculation can reproduce all the physical arguments introduced so cleverly by
CB \cite{cates}.

\textbf{Acknowledgements}

I would like to thank Yohannes Shiferaw for a useful discussion as well as for
pointing out reference \cite{cates} to me. This work is supported by a grant
from the US Department of Energy (DOE) number DE-G02-98ER45686.

\appendix
\section*{}

Here we give some of the intermediate steps leading to Eqs. (\ref{fav}) and
(\ref{nF}). To evaluate the expectation value of the last term in $H_{n}$ it
is most useful to write:
\begin{eqnarray}
&&\left\langle \exp i{\mathbf  k\cdot}\left(  {\mathbf  R}_{a}(u){\mathbf  -R}%
_{b}(u^{\prime})\right)  \right\rangle _{h_{n}} =\int{\mathcal D}%
{\mathbf  R}_{1}\cdots{\mathcal D}{\mathbf  R}_{n}\exp\left(  \sum_{c}\int_{0}%
^{L}dv\ {\mathbf  V}_{c}(v){\mathbf  \cdot R}_{c}(v)\right. \nonumber\\
&& \left.  -\frac12\sum_{cd}\int_{0}^{L}dv\int_{0}^{L}dv^{\prime}%
\ {\mathbf  R}_{c}(v)\ g_{cd}^{-1}(v-v^{\prime})\ {\mathbf  R}_{d}(v^{\prime
})\right) \nonumber\\
&& \times\left(  \int{\mathcal D}{\mathbf  R}_{1}\cdots{\mathcal D}{\mathbf  R}%
_{n}\exp\left(  -\frac12\sum_{cd}\int_{0}^{L}dv\int_{0}^{L}dv^{\prime
}\ {\mathbf  R}_{c}(v)\ g_{cd}^{-1}(v-v^{\prime})\ {\mathbf  R}_{d}(v^{\prime
})\right)  \right)  ^{-1}\nonumber\\
&& =\exp\left(  \frac12\sum_{cd}\int dv\int dv^{\prime}{\mathbf  V}_{c}%
(v)g_{cd}(v-v^{\prime}){\mathbf  V}_{d}(v^{\prime})\right) \nonumber\\
&& =\exp\left(  -\frac12{\mathbf  k}^{2}(g_{aa}(0)+g_{bb}(0)-2g_{ab}(u-u^{\prime
})\right)  ,
\end{eqnarray}
where
\begin{equation}
{\mathbf  V}_{c}(v)=i\ {\mathbf  k\ (\delta}_{c,a}\delta(v-u)-{\mathbf  \delta
}_{c,b}\delta(v-u^{\prime})).
\end{equation}
Next we show how to evaluate other contribution to the free energy:
\begin{eqnarray}
&&\sum_{\omega}\beta G_{aa}(\omega)  =\sum_{\omega\neq0}\frac1{M\omega
^{2}+\lambda}+\frac{\lambda_{1}+2s}{(\lambda_{1}+s)^{2}} \nonumber\\
&& =\frac L{2\sqrt{M\lambda}}\coth\frac L2\sqrt{\frac\lambda M}-\frac
1\lambda+\frac{\lambda_{1}+2s}{(\lambda_{1}+s)^{2}}.
\end{eqnarray}
Also
\begin{equation}
-\frac1{2n}\sum_{ab}p_{ab}\beta G_{ab}(\omega=0)=-\frac1{2n}\mathrm{Tr\ }%
{\mathbf  p\ G}(0)=\frac{\lambda(\lambda_{1}+2s)}{2(\lambda_{1}+s)^{2}}-\frac12,
\end{equation}
and
\begin{eqnarray}
&&-\frac12\sum_{\omega}\mathrm{tr}\ \ln\beta\ {\mathbf  G}(\omega) =\frac
12\sum_{\omega}\mathrm{tr}\ \ln\ \left(  \beta^{-1}{\mathbf  G}^{-1}%
(\omega)\right) \nonumber\\
&& =\frac n2\sum_{\omega}\ln(M\omega^{2}+\lambda)-\frac n2\ln\lambda
+\frac12\mathrm{tr}\ \ln\ \left(  \beta^{-1}{\mathbf  G}^{-1}(0)\right)  ,
\end{eqnarray}
but
\begin{equation}
\frac n2\sum_{\omega}\ln(M\omega^{2}+\lambda)=n\ \ln\left(  2\sinh\ \frac
L2\sqrt{\frac\lambda M}\right)  +n\ {\mathrm const.},%
\end{equation}
(see e.g. \cite{gr} p. 44, Eq. 1.431.2). The constant term (which is infinite)
is eliminated by the normalization of the functional integral, and in any case
does not depend on $\lambda$. Also
\begin{eqnarray}
&&\mathrm{tr}\ \ln\ \left(  \beta^{-1}{\mathbf  G}^{-1}(0)\right) 
=\mathrm{tr}\ \ln\left(
\begin{array}
[c]{cccc}%
\lambda_{1} & -s & \cdots & -s\\
-s & \lambda_{1} & \ddots & \vdots\\
\vdots & \ddots & \ddots & -s\\
-s & \cdots & -s & \lambda_{1}%
\end{array}
\right) \nonumber\\
&& =n\ln\lambda_{1}+n\ln\left(  1+\frac s{\lambda_{1}}\right)  -n\frac
s{\lambda_{1}+s}+o(n^{2}).
\end{eqnarray}

For the case of 1-step RSB we have to calculate the propagator by inverting
Parisi type matrices. It is helpful to use formulas found in an Appendix of
\cite{mp}. We find
\begin{eqnarray}
&&\beta G_{aa}(\omega =0)=\frac1{x_{c}(\lambda_{1}+s_{1}-\Sigma)}+\left(
1-\frac1{x_{c}}\right)  \frac1{\lambda_{1}+s_{1}}+\frac{s_{0}}{(\lambda
_{1}+s_{1}-\Sigma)^{2}},\\
&&\beta G(\omega =0,x)=\frac{s_{0}}{(\lambda_{1}+s_{1}-\Sigma)^{2}%
},\ \ \ x<x_{c},\\
&&\beta G(\omega =0,x)=\frac1{x_{c}(\lambda_{1}+s_{1}-\Sigma)}-\frac1{x_{c}%
}\frac1{\lambda_{1}+s_{1}}+\frac{s_{0}}{(\lambda_{1}+s_{1}-\Sigma)^{2}%
},\ \ \ x>x_{c},\\
&&\beta G_{ab}(\omega \neq0)=\frac1{M\omega^{2}+\lambda}\delta_{ab}\ ,
\end{eqnarray}
where x is Parisi's index on the interval [0,1].

Next we show how various other contribution to the free energy become in the
RSB case:
\begin{equation}
\frac1n\sum_{a}\sum_{\omega}\beta G_{aa}(\omega)=\sum_{\omega\neq0}%
\frac1{M\omega^{2}+\lambda}+\frac1n\sum_{a}\sum_{\omega}\beta G_{aa}%
(\omega=0),
\end{equation}%
\begin{eqnarray}
-\frac1{2n}\sum_{ab}p_{ab}\beta G_{ab}(\omega =0)&=&-\frac1{2n}\mathrm{Tr\ }%
{\mathbf  p\ G}(0) \nonumber\\
& =&\frac\lambda2\beta G_{aa}(\omega=0)-\frac12,
\end{eqnarray}
and finally
\begin{eqnarray}
&&-\frac1{2n}\sum_{\omega}\mathrm{tr}\ \ln\beta\ {\mathbf  G}(\omega) 
=\frac1{2n}\sum_{\omega}\mathrm{tr}\ \ln\ \left(  \beta^{-1}{\mathbf  G}%
^{-1}(\omega)\right) \nonumber\\
&& =\frac12\sum_{\omega}\ln(M\omega^{2}+\lambda)-\frac12\ln\lambda
+\frac1{2n}\mathrm{tr}\ \ln\ \left(  \beta^{-1}{\mathbf  G}^{-1}(0)\right)
\nonumber\\
&& =\ln\left(  2\sinh\ \frac L2\sqrt{\frac\lambda M}\right)  +\frac
12\ln\frac{\lambda_{1}}\lambda+\frac12\ln\left(  1+\frac{s_{1}-\Sigma}%
{\lambda_{1}}\right) \nonumber\\
&& \ -\frac12\left(  1-\frac1{x_{c}}\right)  \ln\left(  1-\frac\Sigma
{\lambda_{1}+s_{1}}\right)  -\frac12\frac{s_{0}}{\lambda_{1}+s_{1}-\Sigma
}+{\mathrm const}.+o(n).
\end{eqnarray}
The coefficient $a_{1}$ in the exponential is given as before by
\begin{equation}
a_{1}=\frac1L\sum_{\omega\neq0}G_{aa}(\omega)\left(  1-e^{i\omega z}\right)  ,
\end{equation}
and $a_{2}$ becomes
\begin{eqnarray}
a_{2}(x)  & =\frac1L\left(  G_{aa}(\omega=0)-G(\omega=0,x)\right)
+\frac1L\sum_{\omega\neq0}G_{aa}(\omega) \nonumber\\
& =a_{2l},\ \ \ \ \ x<x_{c}\\
& =a_{2b},\ \ \ \ x>x,.
\end{eqnarray}
and the explicit expressions for $a_{1}$, $a_{2l}$ and $a_{2b}$ are given in
Eqs. (\ref{a1}), (\ref{a2lf}) and (\ref{a2bf}) respectively. Notice also that
\begin{eqnarray}
\sum_{a\neq b}\exp(-{\mathbf  k}^{2}a_{2}(x)) &=&-\int_{0}^{1}dx\exp
(-{\mathbf  k}^{2}a_{2}(x))\nonumber\\
&=&-x_{c}\exp(-{\mathbf  k}^{2}a_{2l})-(1-x_{c})\exp(-{\mathbf  k}^{2}a_{2b}).
\end{eqnarray}

\end{document}